\documentclass[aip,reprint,eqseqnum,bibnotes,showpacs,superscriptaddress]{revtex4-1}
\usepackage{amsmath,amsthm,amsfonts,amssymb,bm}
\usepackage{graphicx}
\usepackage{color}
\usepackage{natbib}
\usepackage{mciteplus}

\newcommand{\be}{\begin{equation}}
\newcommand{\ee}{\end{equation}}
\newcommand{\bea}{\begin{eqnarray}}
\newcommand{\eaa}{\end{eqnarray}}

\newcommand{\tw}{t_w}

\newcommand{\gdot}{\dot{\gamma}}
\newcommand{\gdotbar}{\overline{\dot{\gamma}}}
\newcommand{\gbar}{\overline{\gamma}}

\newcommand{\Sigmay}{\Sigma_{\rm y}}
\newcommand{\sigmay}{\sigma_{\rm y}}

\newcommand{\kT}{k_{\rm B}T}
\newcommand{\xg}{x_{\rm g}}

\usepackage{graphicx,amsmath,color,amsfonts}

\begin{document}
 
\title{Shear banding in soft glassy materials}

\author{S. M. Fielding} 

\affiliation{Department of Physics, Durham University, Science
  Laboratories, South Road, Durham, DH1 3LE, UK}

\date{\today}
\begin{abstract} 
  Many soft materials, including microgels, dense colloidal emulsions,
  star polymers, dense packings of multilamellar vesicles, and
  textured morphologies of liquid crystals, share the basic ``glassy''
  features of structural disorder and metastability.  These in turn
  give rise to several notable features in the low frequency shear
  rheology (deformation and flow properties) of these materials: in
  particular, the existence of a yield stress below which the material
  behaves like a solid, and above which it flows like a liquid.  In
  the last decade, intense experimental activity has also revealed
  that these materials often display a phenomenon known as shear
  banding, in which the flow profile across the shear cell exhibits
  macroscopic bands of different viscosity.  Two distinct classes of
  yield stress fluid have been identified: those in which the shear
  bands apparently persist permanently (for as long as the flow
  remains applied), and those in which banding arises only transiently
  during a process in which a steady flowing state is established out
  of an initial rest state (for example, in a shear startup or step
  stress experiment).  After surveying the motivating experimental
  data, we describe recent progress in addressing it theoretically,
  using the soft glassy rheology model and a simple fluidity model. We
  also briefly place these theoretical approaches in the context of
  others in the literature, including elasto-plastic models, shear
  transformation zone theories, and molecular dynamics simulations. We
  discuss finally some challenges that remain open to theory and
  experiment alike.

\end{abstract}
\maketitle


\newif\iffigures
\figurestrue 

\section{Introduction}
\label{sec:intro}

Many soft materials, including microgels, dense colloidal emulsions,
star polymers, dense packings of multilamellar vesicles, and textured
morphologies of liquid crystals, share several notable features in
their rheological (deformation and flow) properties. In a steady shear
experiment, the `flow curve' relation $\Sigma(\gdot)$ between shear
stress $\Sigma$ and shear rate $\gdot$ is
often~\cite{Holdsworth93,BarHutWal89,Dickinson92} fit to the form
$\Sigma-\Sigmay\propto\gdot^n$, either with a non-zero apparent yield
stress $\Sigmay\neq 0$, or with `power-law fluid' behaviour for
$\Sigmay=0$.  Likewise their viscoelastic spectra, measured in a small
amplitude oscillatory shear deformation, exhibit a characteristically
flat power-law form over several decades of frequency, even at the
lowest frequencies accessible
experimentally~\cite{KetPruGra88,KhaSchneArm88,MasoWeit95b,MBW95,ISI:A1996TT76300005}.
The same materials often also exhibit rheological
ageing~\cite{ISI:000083042700015,ISI:000165556600044,ISI:000168623500069,ISI:000177009600030,ISI:000180338300019,ISI:000181015900058,mann1,mann2,mann3},
in which a sample slowly evolves towards an ever-more solid-like state
as a function of its own age: {\it i.e.,} of the time elapsed since
sample preparation.

The widespread observation of these unifying signatures suggests a
common cause. Indeed, all these materials share the basic features of
structural disorder and metastability. In a dense packing of emulsion
droplets, for example, large energy barriers $E\gg \kT$ associated
with stretching the interfaces between the droplets impede
rearrangements of the droplets relative to one another. In consequence
the system may become arrested in disordered, metastable droplet
configurations on very long timescales, even if the state of lowest
free energy might in principle be ordered. These materials are
therefore non-ergodic, and in this sense can be viewed as ``glassy''.
The term ``soft glassy materials'' (SGMs)~\cite{ISI:A1997WM06400048}
has been coined to describe to them.

Beyond the rheological features described above, which have been
discussed in detail in previous papers, an accumulating body of
experimental data further indicates that heterogeneous ``shear
banded'' flow states often arise when these materials are subject to
an imposed shear
flow~\cite{PhysRevLett.88.175501,martin-sm-8-6940-2012,paredes-jpm-23--2011,requested1,requested2,requested3,requested4,rogers-prl-100--2008,RT114,ISI:000294447600069,ISI:000280140800011}.
A rich interplay can then take place between this flow heterogeneity
that forms on a macroscopic lengthscale, and the material's underlying
ageing dynamics. This has a major influence on bulk rheological
properties, and so also potentially on any industrial application in
which these materials are subject to flow (whether during processing
or directly in use), and/or that involve a long shelf life before use.

The aim of this key issues article is to review recent theoretical
progress~\cite{Moorcroftetal2012,ISI:000286879900011,ISI:000266798200007}
in modeling these shear banded flows of densely packed soft glassy
materials, as well as of other yield stress fluids such as gels
comprising low-density space-spanning networks of attractive colloidal
particles.  It does so from the particular subjective viewpoint of the
``soft glassy rheology'' (SGR) model~\cite{ISI:A1997WM06400048}, of
which this author has the most direct experience. In some places we
also supplement our SGR results with studies of a simple ``fluidity''
model~\cite{ISI:000286879900011}: partly to demonstrate that the
phenomena we address are not model-specific, and partly because
fluidity models are often more convenient to study numerically.

In a review of this relative brevity it is impossible to describe
exhaustively all other theoretical approaches in the literature.
Nonetheless, we shall attempt briefly to place our own findings in the
context of some other approaches: including shear transformation zone
(STZ)
theories~\cite{manning-pre-79--2009,manning-pre-76--2007,falk-arcmpv2-2-353-2011},
models of coupled elasto-plastic dynamical
events~\cite{picard-epje-15-371-2004,picard-pre-71--2005,mansard-sm-7-5524-2011,martens-sm-8-4197-2012,jagla-jsme---2010,bocquet-prl-103--2009,jagla-pre-76--2007},
fluidity models besides the one used
here~\cite{picard-pre-66--2002,coussot-jr-46-573-2002}, and molecular
dynamics
simulations~\cite{varnik-prl-90--2003,varnik-jcp-120-2788-2004,xu-pre-73--2006,chaudhuri-pre-85--2012}.
The reader is encouraged to explore the references provided in these
areas.

The manuscript is structured as follows. In Sec.~\ref{sec:experiment}
we survey the experimental evidence for shear banding in soft glassy
materials. In Sec.~\ref{sec:models} we describe the SGR and fluidity
models.  In Secs.~\ref{sec:transient} and~\ref{sec:permanent} we
review recent results for shear banding in soft glassy materials,
obtained within these models.  Finally in Sec.~\ref{sec:conclusion} we
give conclusions and perspectives for further study.

\section{Experimental motivation}
\label{sec:experiment}

The rheological properties of yield stress fluids (YSFs) have been
intensively investigated during the last decade. In the vicinity of
the yield stress, two apparently distinct classes of rheological
behaviour have been identified.  The
first
~\cite{paredes-jpm-23--2011,ragouilliaux-pre-76--2007,divoux-prl-104--2010,divoux-prl-110--2013,divoux-sm-7-9335-2011,ISI:000294447600069,moller-e-87--2009,divoux-prl-110--2013,requestedN}
is characterised by a continuous transition between solid-like and
liquid-like behaviour as the stress increases above $\Sigmay$, with
the shear rate increasing smoothly from zero as a function of
$\Sigma-\Sigmay$. Associated with this smooth transition in a
controlled stress protocol is the observation in slow up/down
controlled shear rate sweeps of only minor hysteresis
effects~\cite{divoux-prl-104--2010,mann4}.  Furthermore, under conditions of
a constant imposed shear rate, the steady flowing state is one of
homogeneous shear for all values of the imposed shear rate, however
small.  Materials that exhibit these characteristics have come to be
termed ``simple YSFs'' in the literature.  Examples include emulsions
and carbopol microgels.  They typically have predominantly repulsive
interactions between their constituent mesoscopic substructures.

A second, contrasting class of rheological behaviour has been
identified in YSFs that have attractive interactions between the
constituent particles.  These materials display a discontinuous
transition from solid-like to liquid-like behaviour on increasing the
imposed stress above a critical threshold value, with the shear rate
jumping discontinuously from zero just below the threshold stress to a
finite value $\gdot_{\rm c}$ just above it.  Equivalently, the
viscosity jumps discontinuously from being effectively infinite (at
long times) just below threshold to being finite just above it: an
effect referred to as ``viscosity
bifurcation''~\cite{PhysRevLett.88.175501,coussot-jr-46-573-2002,ragouilliaux-pre-76--2007,da-pre-66--2002}.
The apparently forbidden window of shear rates $\gdot=0\to\gdot_{\rm
  c}$ associated with this jump is then found to correspond, under
conditions of a constant imposed shear rate in this range
$\gdot=0\to\gdot_{\rm c}$, to the long time response of the material
being shear
banded~\cite{PhysRevLett.88.175501,martin-sm-8-6940-2012,paredes-jpm-23--2011,rogers-prl-100--2008,RT114}.

Associated with this observation of shear banding under conditions of
a constant imposed shear rate is the presence in slow up/down shear
rate sweeps of strong hysteresis effects
~\cite{divoux-prl-110--2013,rogers-jr-54-133-2010,rogers-prl-100--2008,divoux-prl-110--2013,mann5}.
Because of this pronounced hysteresis, materials in this second
category are often referred to in the literature as ``thixotropic
YSFs'', in an attempt to distinguish them from the ``simple YSFs''
discussed above.  However even simple YSFs are sometimes seen to show
age dependence and shear-rejuvenation ({\it i.e.}, thixotropy, using
the definition in Ref. of~\cite{mewis} ``the continuous decrease of
viscosity with time when flow is applied to a sample that has been
previously at rest and the subsequent recovery of viscosity in time
when the flow is discontinued'') in their time-dependent flow
behaviour~\cite{divoux-sm-7-9335-2011}.  We therefore prefer instead
to label these two classes of material as ``simple YSFs'' and
``viscosity-bifurcating YSFs'' respectively, and accordingly shall do so
throughout the manuscript.


Between these two classes of material, then, only
viscosity-bifurcating YSFs appear to display shear banding as their
long-time, ``permanent'' response to a steady imposed shear rate.
This permanent banding is then strongly reminiscent of the steady
state shear banding that has been intensively investigated in ergodic
complex fluids such as
polymers~\cite{ISI:000263272300011,ISI:000281474500006,ISI:000254645200053,tapadiawang06}
and wormlike
micelles~\cite{ISI:000253331200055,B900948E,Helgeson:09,hu:379,PhysRevE.63.022501,lerouge-sm-4-1808-2008,decruppe2006,britton-prl-78-4930-1997,britton-epjb-7-237-1999,MairCall96,miller2007,salmon-prl-90--2003}.
In particular, macroscopic bands of unequal shear rates $\gdot_{\rm
  low},\gdot_{\rm high}$ coexist at a common value of the shear
stress, with the relative volume fraction of the bands controlled by a
lever rule~\cite{RT114}. The associated signature in the material's
bulk rheology is a characteristic plateau in the composite flow curve
$\Sigma(\gdotbar)$ (where $\gdotbar$ now denotes the shear rate
applied to the sample as a whole, averaged across the bands).
Distinct from conventional ergodic fluids, however, in these YSFs the
high viscosity band is effectively unsheared, $\gdot_{\rm low}\approx
0$, and displays ageing
dynamics~\cite{rogers-prl-100--2008,rogers-jr-54-133-2010,RT114,mann5}.

In ergodic complex fluids, the criterion for the formation of steady
state shear bands is widely known: that the constitutive relation
$\Sigma(\gdot)$ between shear stress and shear rate for an underlying
base state of homogeneous shear flow should have a region of negative
slope, $d\Sigma/d\gdot<0$. Theories based on this concept of a
non-monotonic constitutive curve have likewise been put forward to
explain permanent (though ageing) shear bands in viscosity-bifurcating
YSFs~\cite{picard-pre-66--2002,coussot-jr-46-573-2002,mansard-sm-7-5524-2011,martens-sm-8-4197-2012,jagla-pre-76--2007,manning-pre-79--2009},
as we shall describe in Sec.~\ref{sec:permanent} below. Other
studies~\cite{varnik-prl-90--2003,varnik-jcp-120-2788-2004,xu-pre-73--2006,chaudhuri-pre-85--2012}
suggest alternatively the existence of a static yield stress
$\Sigma_{\rm Ys}$ (in well rested samples) below which there exists a
branch of zero flow states, with $\Sigma_{\rm Ys}$ exceeding the
dynamical yield stress $\Sigma_{\rm Y}$ as measured in a protocol that
instead sweeps the shear rate down towards zero. (That dynamical
branch is itself purely monotonic.)  This gives rise to a downwards
step discontinuity (non-monotonicity) in the material's constitutive
properties at $\gdot=0$ (see Fig.~\ref{fig:sketch} below) and again
allows a coexistence of an unsheared band with a flowing one. We
return in Sec.~\ref{sec:conclusion} to discuss these studies in
relation to those of Sec.~\ref{sec:permanent}.

In contrast, simple YSFs do not form permanent shear bands under a
constant imposed shear rate.  However it has recently become apparent
that shear banding can arise quite generically in {\em time-dependent}
flow
protocols~\cite{ISI:000286879900011,Moorcroftetal2012,Moorcroftthesis,MoorcroftPreprint},
even in materials that have a monotonic $\Sigma(\gdot)$ and therefore
lack the possibility of permanent banding under a steady imposed
shear. (This statement in fact applies to all complex fluids, and not
just the soft glassy ones of interest here. For a recent study of this
issue in polymeric and wormlike micellar fluids, see
Ref.~\cite{MoorcroftPreprint}) In a shear startup protocol, for
example, the (almost ubiquitously observed) signature of an overshoot
in the stress startup curve $\Sigma(t)$ is thought to be generically
associated with the formation of shear bands.  Once formed, these
bands may persist only transiently, or may remain to steady state,
according to whether the material's underlying constitutive curve
$\Sigma(\gdot)$ is monotonic or non-monotonic.  Likewise in a step
stress experiment, the presence of simultaneously upward slope and
upward curvature in the material's time-differentiated creep response
curve $\gdot(t)$ is thought to give rise generically to the formation
of shear bands~\cite{Moorcroftetal2012}.

Consistent with these predictions, experiments have indeed revealed
shear banding in YSFs during the time-dependent protocols of step
stress and shear startup: both in viscosity-bifurcating
YSFs~\cite{martin-sm-8-6940-2012}, which can then remain shear banded
even once the system attains its final steady flowing state, {\em and}
in simple YSFs~\cite{ISI:000294447600069}
which by definition recover a homogeneous flow in steady state. We
shall now briefly review these experiments, discussing in turn the
step stress and shear startup protocols.

Following the imposition of a step stress, the most commonly reported
rheological response function is the creep curve $\gamma(t)$ (or its
time-differential $\gdot(t)$), which describes how the material creeps
and (perhaps) eventually flows in response to the applied load.
Experimentally, these curves show strikingly similar features across a
range of YSFs as diverse as ketchup~\cite{caton},
mustard~\cite{caton}, mayonnaise~\cite{caton}, hair gel~\cite{caton},
carbopol~\cite{caton,ISI:000294447600069}, a hard sphere colloidal
glass~\cite{siebenbuerger-prl-108--2012}, carbon
black~\cite{sprakel-prl-106--2011,ISI:000280140800011},
thermoreversible gels~\cite{mann6} and a lyotropic hexagonal columnar
phase~\cite{bauer-prl-97--2006}, as we shall now describe.

For an applied stress $\Sigma<\Sigmay$, one typically observes a
process of very slow creep in which the shear rate progressively tends
towards zero at long times, often in the form of a power law
$\gdot\sim t^{-\beta}$. In contrast, for an imposed stress just above
$\Sigma_{\rm Y}$ the strain response typically shows a succession of
several distinct regimes. Initially, the material creeps as though at
an imposed stress just below yield, with $\gdot\sim t^{-\beta}$.
(This phenomenon is often referred to as Andrade creep, following
original observations by that author in 1910 of {\rm tensile} creep in
metallic wires~\cite{andrade-1910}.)  This regime of slow creep then
terminates at some fluidisation time $\tau_{\rm f}=\tau_{\rm
  f}(\Sigma-\Sigmay)$, when the shear rate suddenly curves upwards and
dramatically increases before finally curving downwards onto a steady
flowing state of time-independent $\gdot_{\rm ss}=\gdot_{\rm
  ss}(\Sigma-\Sigmay)$. For stress values approaching the yield from
above, $\Sigma\to\Sigmay^+$, one typically finds $\gdot_{\rm ss}\to 0$
and $\tau_{\rm f}\to\infty$ for a simple
YSF~\cite{ISI:000294447600069}. For viscosity-bifurcating YSFs
$\gdot_{\rm ss}$ and $\tau_{\rm f}$ should both remain finite as
$\Sigma\to\Sigmay^+$ before $\gdot_{\rm ss}$ jumps discontinuously to
zero below yield.  (The actual functional form of the fluidization
time versus stress in the vicinity of yield is complicated, varying
from material to material and also depending on wall effects.)

In Ref.~\cite{ISI:000294447600069} these bulk creep curves are
measured in tandem with spatially resolved flow profiles across the
cell following the imposition of a step stress in carbopol microgel (a
simple YSF).  During the initial phase of slow creep, the shear rate
profile remains homogeneous.  Subsequently, at the onset of
fluidisation, strong wall slip arises that quickly gives way to bulk
shear banding as the shear rate increases strongly.  The bands then
decay as the shear rate curves downwards onto its final steady state
value.  The eventual steady flowing state is then homogeneous,
consistent with carbopol being a simple YSF.  Slow creep followed by
fluidisation during which shear localisation occurred was also
reported in a viscosity-bifurcating YSF in
Ref.~\cite{ISI:000280140800011}.

In a shear startup experiment, a previously well rested sample is
subject for all times $t>\tw$ to shear of constant rate $\gdot$,
eventually leading to a steady flowing state in the limit
$t\to\infty$. Here $\tw$ denotes the time elapsed since sample
preparation before the flow commenced.
Refs.~\cite{divoux-sm-7-9335-2011,divoux-prl-104--2010} report startup
experiments on carbopol, again with spatially resolved velocimetry of
the flow profiles across the shear cell reported in tandem with bulk
rheological measurements. At early times, the stress startup signal
$\Sigma(t)$ grows linearly with strain, $\Sigma=G\gamma=G\gdot t$, and
the shear field remains uniform across the cell.  At longer times the
stress shows a strong overshoot, the height of which depends strongly
on the sample age $\tw$ before the flow commenced. The stress then
falls from this overshoot as it descends to a final steady state. As
it does so, pronounced wall slip followed by macroscopic shear banding
are observed in the flow profiles.  At longer times the bands decay to
again leave a homogeneous shear profile in steady state, consistent
with carbopol being a simple YSF.

Shear band formation triggered by stress overshoot in shear startup
was also reported in a viscosity-bifurcating YSF in
Ref.~\cite{martin-sm-8-6940-2012}.  In that case the bands persist to
steady state, consistent with the more complex rheology of viscosity
bifurcating YSFs.

\section{SGR and fluidity models}
\label{sec:models}

Having summarised the experimental phenomenology of shear banding in
soft glassy materials, we now discuss the two models that we shall use
to address it theoretically: the SGR model and a simple fluidity
model. We introduce them here in their original form as first put
forward in Refs.~\cite{ISI:A1997WM06400048,ISI:000286879900011}: each
with a monotonic constitutive curve, and so capable of addressing only
simple YSFs.  In Sec.~\ref{sec:permanent} below we shall introduce a
simple modification to the SGR model that captures a non-monotonic
constitutive curve and thereby addresses the permanent shear banding
seen in viscosity-bifurcating YSFs, with layer normals in the
flow-gradient direction. A similar modification to fluidity models,
with permanent banding, was discussed earlier in
Ref.~\cite{picard-pre-66--2002}. We do not address in this work the
possibility of transverse banding in which heterogeneity develops in
the vorticity direction, which in a glassy system might be expected to
be associated with a discontinuous shear thickening
transition~\cite{vorticity}.

\subsection{Soft glassy rheology model}
\label{sec:SGR}

The soft glassy rheology (SGR) model~\cite{ISI:A1997WM06400048} is
based on Bouchaud's model of glassy dynamics\cite{Bouchaud92}. It
considers an ensemble of elements undergoing independent activated
hopping dynamics among a (free) energy landscape of traps. In the
context of a soft glassy material, each element is taken to represent
a mesoscopic cluster of, say, a few tens of emulsion droplets. For any
such element it is assumed possible to identify local continuum
variables of shear strain $l$ and shear stress $kl$, which describe
the cluster's state of local elastic deformation relative to a state
of locally undeformed equilibrium.  Rheology is incorporated by
assuming that, between hops, the strain of each element affinely
follows the macroscopic flow field to which that region of material is
subject: $\dot{l}=\gdot$.  The stress of the sample as a whole is
defined as the average over the local elemental ones: $\sigma=\langle
kl\rangle$.

The hopping of any element out of one trap and into another is then
identified with a local yielding event in which a cluster of droplets
suddenly rearranges into a new configuration locally.  In doing so, it
is assumed to select a new trap depth at random from a prior
distribution $\rho(E)\sim\exp(-E/\xg)$, and to reset its local strain
$l$ to zero.  It is these yielding events, then, that confer
rheological stress relaxation.

Hops are taken to be dynamically activated, such that an element in a
trap of depth $E$ and with local shear strain $l$ has a probability
per unit time of yielding given by
$\tau^{-1}(E,l)=\tau_0^{-1}\exp\left[-(E-\tfrac{1}{2}kl^2)/x\right]$.
In this way, the element's stored elastic energy $\tfrac{1}{2}kl^2$ at
any instant offsets the trap-depth $E$, leading to a reduced local
barrier to rearrangement $E-\tfrac{1}{2}kl^2$. This leads to
rheological shear thinning in the sample as a whole.
Because the typical energy barrier $E\gg \kT$, the parameter $x$ is
not the true thermodynamic temperature but rather an effective noise
temperature that models in a mean field way coupling with other
yielding events elsewhere in the sample.

Combined with the exponential prior $\rho(E)$, the exponential
activation factor just described ensures (in the absence of flow at
least) a glass transition at a noise temperature $x=\xg$.  For noise
temperatures $x<\xg$, the model shows a yield stress $\sigmay$ that
initially rises linearly with $\xg-x$ just below the glass point.  In
the absence of an imposed flow (or more generally for sample stresses
$\sigma<\sigmay$), rheological ageing occurs: following sample
preparation at time $t=0$ by means of a sudden quench from a high
initial noise temperature to a value $x<\xg$, the system progressively
evolves into ever deeper traps as a function of time since
preparation. In rheological terms, this corresponds to a growing
stress relaxation time $\langle \tau\rangle\sim t$, and so to ever
more solid-like response as a function of the sample age.  An imposed
shear of constant rate $\gdot$ can however arrest ageing and
rejuvenate the sample to a steady flowing state of effective age
$\langle \tau\rangle\sim 1/\gdot$

According to the dynamics just described, then, the probability
$P(E,l,t)$ for an element to be in a trap of depth $E$ and with local shear
strain $l$ obeys
\be
\label{eqn:master}
\dot{P}(E,l,t)+\gdot\frac{\partial P}{\partial l} = -\frac{1}{\tau(E,l)}P+Y(t)\rho(E)\delta(l).
\ee
The convected derivative on the left hand side describes affine
loading of each element by shear. The first and second terms on the
right hand side describe hops out of and into traps respectively, with
an ensemble average hopping rate
\be
Y(t)=\int dE \int dl \frac{1}{\tau(E,l)}P(E,l,t).
\ee
The macroscopic stress
\be
\sigma(t)=\int dE \int dl\; kl P(E,l,t).
\ee
Throughout we use units in which $\tau_0=1$, $k=1$ and $\xg=1$.

Numerical results for the SGR model's flow
curves~\cite{ISI:000074893400095} $\sigma(\gdot)$ are shown in
Fig.~\ref{fig:SGRconstitutive} for a range of noise temperatures $x$.
For $x\ge 2$ these display Newtonian response with $\sigma\sim \gdot$;
for $1<x<2$ we see power-law fluid behaviour with
$\sigma\sim\gdot^{x-1}$; and for $x<1$ a non-zero yield stress with
$\sigma-\sigmay(x)\sim \gdot^{1-x}$.

\iffigures
\begin{figure}[tbp]
  \centering
  \includegraphics[width=0.5\textwidth]{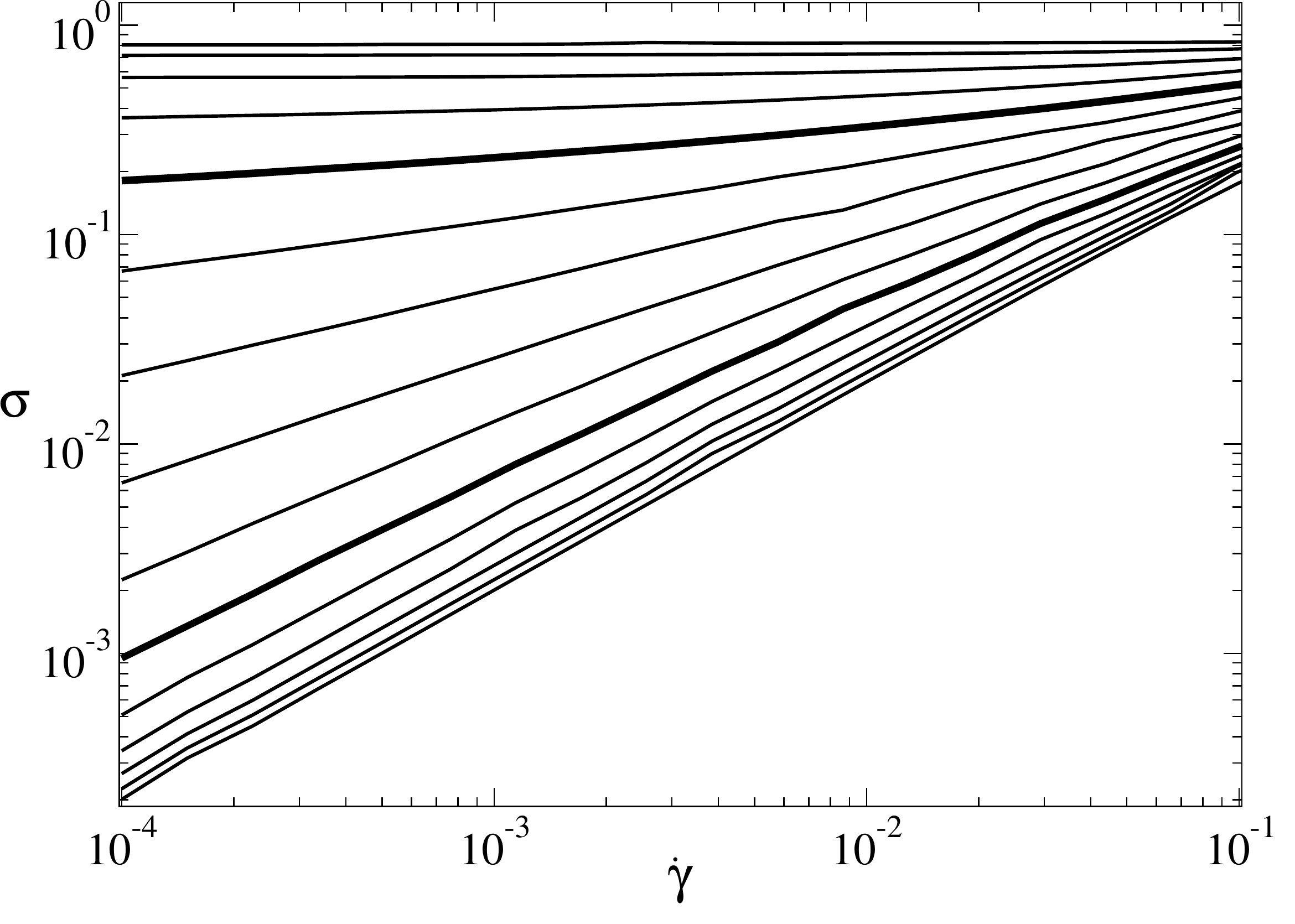}
  \caption{Constitutive curves of the SGR model for noise temperatures $x=0.2,0.4\cdots 3.0$ (curves downward). Bold lines highlight $x=1$ and $x=2$.}
  \label{fig:SGRconstitutive}
\end{figure}
\fi

So far, we have described the SGR model in its original form as
introduced in Ref.~\cite{ISI:A1997WM06400048}. In this form, the model
contains no spatial information about the location of any element and,
as such, is obviously incapable of addressing shear banded flows in
which the flow state of the material varies in the flow-gradient
direction $y$. In Ref.~\cite{ISI:000266798200007}, therefore, we
extended the model to allow spatial variations in this dimension $y$
(with translational invariance still assumed in the flow direction $x$
and vorticity direction $z$). To do so we discretized the $y$
coordinate into $i=1 \cdots n$ streamlines of equal spacing $L_y/n$,
giving an overall sample thickness $L_y$.  For convenience we adopted
periodic boundary conditions such that $i=1,n$ are also neighbours.
Each streamline is then assigned a separate ensemble of $j=1\cdots m$
SGR elements, with a streamline shear stress $\sigma_i=(k/m)\sum_j
l_{ij}$.

In the creeping flow conditions of interest here, the force balance
condition decrees that the shear stress is uniform across all
streamlines, $\sigma_i(t)=\sigma(t)$.  During intervals in which no
jump occurs anywhere in the system, the material clearly deforms as a
linear elastic solid with $\dot{l}=\gdot$ for every element on all
streamlines: any stress change is uniform across all streamlines,
consistent with force balance. Supposing a hop then occurs at element
$ij$ when its local strain is $l=\ell$.  (Numerically, we handle the
hopping dynamics by a waiting-time Monte Carlo algorithm that
stochastically chooses both the element and time of the next hop.)
This clearly reduces the stress on that streamline, in potential
violation of force balance. By updating all elements on the same
streamline $i$ as $l\to l + \ell/m$, force balance then is restored
across the streamlines, but (incorrectly) with a stress level that has
not been properly reduced by the yielding event.  Further updating all
elements on all streamlines throughout the system as $l \to l -
\ell/mn$ restores the global stress to the properly reduced level.

This algorithm can be thought of as the $\eta\to 0$ limit of a
situation in which a small Newtonian viscosity $\eta$ is present
alongside the elastic stress of the SGR elements, with local strain
rates set to maintain a uniform total stress
$\Sigma(t)=\sigma(y,t)+\eta\gdot(y,t)$ across the sample at all times,
in accordance with the force balance condition.  For the long
timescales $\tau \gg \eta/k$ and low flow rates $\gdot \ll k/\eta$ of
interest here, taking this limit $\eta \to 0$ upfront is an excellent
approximation.

Finally, a small diffusivity of stress~\cite{ISI:000084891700016}
between neighbouring streamlines is needed to ensure that the
interface between any shear bands has a slightly diffuse width, rather
than being an unphysical step discontinuity.  This is incorporated by
further adjusting the strain of three randomly chosen elements on each
adjacent streamline $i\pm 1$ by $\ell w(-1,+2,-1)$, after a hop as
described above on streamline $i$.  For the small values of the
coupling strength $w$ of interest here this mildly changes the model's
flow curves relative to those in Fig.~\ref{fig:SGRconstitutive}, but
without changing their overall shape.

\subsection{Fluidity model}
\label{sec:fluidity}

In the previous subsection we introduced the SGR model, which
considers an ensemble of elastic elements undergoing activated hopping
dynamics among an energy landscape of traps, with the probability
$P(E,l,t)$ of finding an element in a trap of depth $E$ and with a
shear strain $l$ evolving according to Eqn.~\ref{eqn:master}. Out of
this full probability distribution one can then define a hierarchy of
moments
\be
P_{p,q}=\int_{-\infty}^{\infty}dl\int_0^{\infty}dE\,\frac{l^p}{\tau^q}P(E,l,t)\;\;\textrm{for}\;\;p,q=0,1,2\cdots,
\ee
and it is easy to show that these evolve according to
\be
\dot{P}_{p,q}=\gdot pP_{p-1,q}+\frac{\gdot q}{x}P_{p+1,q}-P_{p,q+1}+\frac{1}{1+q/x}P_{0,1}\delta_{p,0}.
\label{eqn:moments}
\ee
To fully solve the SGR model's dynamics in this representation, one would
of course need to evolve this infinite hierarchy of coupled moments.

In this section we introduce a simplified ``fluidity'' description,
motivated by the momentwise representation of the full SGR model just
described~\cite{ISI:000286879900011}, and along the lines of earlier
fluidity models in
Refs.~\cite{picard-pre-66--2002,PhysRevLett.88.175501}. It considers
just two moments: the macroscopic stress $\sigma=P_{1,0}$ and the
average hopping rate $Y=P_{0,1}$. The latter of these is often termed
the material's fluidity, and in fact we further cast it in terms of
the material's overall structural relaxation timescale, denoted
$\tau\equiv 1/Y$.  (The overall $\tau$ defined here is of course
distinct from, though related to, the local elemental ones of the full
SGR model above.) The fluidity model then directly writes down
equations of motion for these two moments, in a form that is loosely
inspired by the SGR dynamics for $\dot{P}_{1,0}$ and $\dot{P}_{0,1}$
as given by Eqn.~\ref{eqn:moments} above, but without attempting self
consistent closure with regards the other moments.

To set up the fluidity model, then, we decompose the total shear stress
in any fluid element into a viscoelastic contribution $\sigma$, and a
small Newtonian contribution:
\be
\Sigma(t)=\sigma(y,t)+\eta\gdot(y,t).
\ee
As in the SGR model above, we shall be interested in the limit in
which the Newtonian contribution acts to ensure force balance on a
short timescale set by $\eta$, but otherwise makes negligible
contribution.

The dynamics of the viscoelastic contribution to the stress is
prescribed by a Maxwell-like model
\be
\partial_t\sigma(y,t)=G\gdot-\frac{\sigma}{\tau},
\ee
describing elastic loading with a modulus $G$, and viscoelastic
relaxation on a timescale $\tau$. The relaxation timescale $\tau$ is
then assigned its own dynamics
\be
\partial_t
\tau(y,t)=1-\frac{\tau}{\tau_0+1/\gdot}.
\label{eqn:tau}
\ee

In the absence of flow this gives rheological ageing with a growing
stress relaxation time $\tau\sim t$: the material evolves towards a
progressively more solid-like and less fluid-like state as a function
of its own age, following sample preparation assumed to take place via
a deep quench at time $t=0$, such that $\tau(y,t=0)=\tau_0$.
Conversely, under an imposed shear at a constant rate $\gdot$, ageing
is cutoff at an effective sample age $\tau=\tau_0+1/\gdot$. The steady
state flow curve is then
\be
\Sigma=G(1+\gdot\tau_0)+\eta\gdot,
\ee
rising monotonically in $\gdot$ beyond a yield stress $\Sigmay=G$.

Finally we add to Eqn.~\ref{eqn:tau} a diffusive term
$l_0^2\partial_y^2\tau$ to give a small coupling between
streamlines~\cite{ISI:000084891700016} and confer on the interface
between bands a slightly diffuse width $O(l_0)$.  Throughout we use
units in which $G=1$, $\tau_0=1$, and the width of the flow domain
$L_y=1$.

In the next two sections in turn we shall consider the predictions of
the SGR and fluidity models for the shear banding behaviour of simple
YSFs and (with a simple modification to account for a non-monotonic
constitutive curve) viscosity-bifurcating YSFs.

\section{Simple yield stress fluids}
\label{sec:transient}

For a complex fluid subject to a steady imposed shear flow, the
criterion for the formation of shear bands that will persist
`permanently' -- {\it i.e.,} for as long as the flow remains applied
-- is well known: that the constitutive relation $\Sigma(\gdot)$
between shear stress $\Sigma$ and shear rate $\gdot$ for an underlying
base state of homogeneous flow has a region of negative slope,
$d\Sigma/d\gdot<0$. This criterion is universal to all complex fluids
and applies not only to the non-ergodic soft glassy materials of
interest here, but also to ergodic fluids such as polymers and
wormlike micellar surfactants.  In the context of a soft glass with a
yield stress, it corresponds to a constitutive curve of the shape in
Fig.~\ref{fig:model1}a) below. In their form as described above,
however, the SGR and fluidity models have monotonic constitutive
curves and are unable to address the permanent banding seen under
conditions of a steady applied flow in viscosity-bifurcating YSFs.
We shall nonetheless return in Sec.~\ref{sec:permanent} below to
discuss a simple modification to the SGR model that does allow
permanent banding.

Besides steady shear, many practical flow situations involve a strong
time dependence, whether perpetually or during the transient process
whereby a steady flowing state is established in an sample that was
previously well rested. Commonly studied protocols include step
stress, step strain, and shear startup.  In recent years a body of
experimental data has accumulated to show that, in many complex
fluids, pronounced shear banding can arise during these time-dependent
flow protocols, even if the eventual steady flowing state is unbanded.
This has been observed in the non-ergodic soft glassy materials of
interest here, as surveyed in Sec.~\ref{sec:experiment} above, as well
as in polymeric fluids. (See Ref.~\cite{MoorcroftPreprint} and
references therein to the experimental polymer literature in this
area.)

Motivated by these observations, in a recent
Letter~\cite{Moorcroftetal2012} we derived criteria for the onset of
linear instability to the formation shear bands in time-dependent
flows, one for each protocol in turn: step stress, shear startup, and
step strain.  Importantly, each criterion depends only on the shape of
the experimentally measured rheological response function for that
protocol, but is otherwise {\em independent} of the constitutive
properties of the particular fluid in question. In this way the
criteria apply universally to all complex fluids and have the same
highly general status as the widely known criterion for permanent
banding in steady shear (of a negatively sloping constitutive curve).

In the next two subsections we discuss the application of these
criteria to the onset of shear banding in time-dependent flows of soft
glassy materials, with supporting numerical evidence provided by
simulating the SGR and fluidity models. A counterpart investigation in
the context of polymeric fluids (polymer solutions, polymer melts and
wormlike micellar surfactants) was recently performed in
Ref.~\cite{MoorcroftPreprint}, with supporting numerical evidence from
the rolie-poly and Giesekus constitutive models.

These predictions for the onset of banding in time-dependent flows are
in fact expected to apply to both simple YSFs and
viscosity-bifurcating YSFs, given that the instability criteria are
universal in the way just described. However the focus in this section
is on simple YSFs, for which they have been investigated most
thoroughly experimentally, and to which the SGR model in its original
form applies. Accordingly, the time-dependent bands predicted in this
section always decay to leave homogeneous flow in steady state,
consistent with the behaviour of a simple YSF. In a viscosity
bifurcating fluid we would expect onset in the same way as predicted
here following a step stress or shear startup, but with the bands
persisting permanently for applied shear rates $\gdot<\gdot_{\rm c}$.

\subsection{Step stress protocol: slow creep and fluidisation}
\label{sec:creep}

Consider an experimental protocol in which a sample is freshly
prepared in a reproducible state at some time $t=0$, for example by
loading into a rheometer and preshearing, then left to age in the
absence of any applied flow or loading during a waiting time $\tw$. At
this time it is suddenly subject to a step shear stress of size
$\Sigma_0$, which is held fixed for all subsequent times:
\be
\Sigma(t)=\Sigma_0\Theta(t-\tw).
\ee

\iffigures
\begin{figure*}[tbp]
  \centering
  \includegraphics[width=1.0\textwidth]{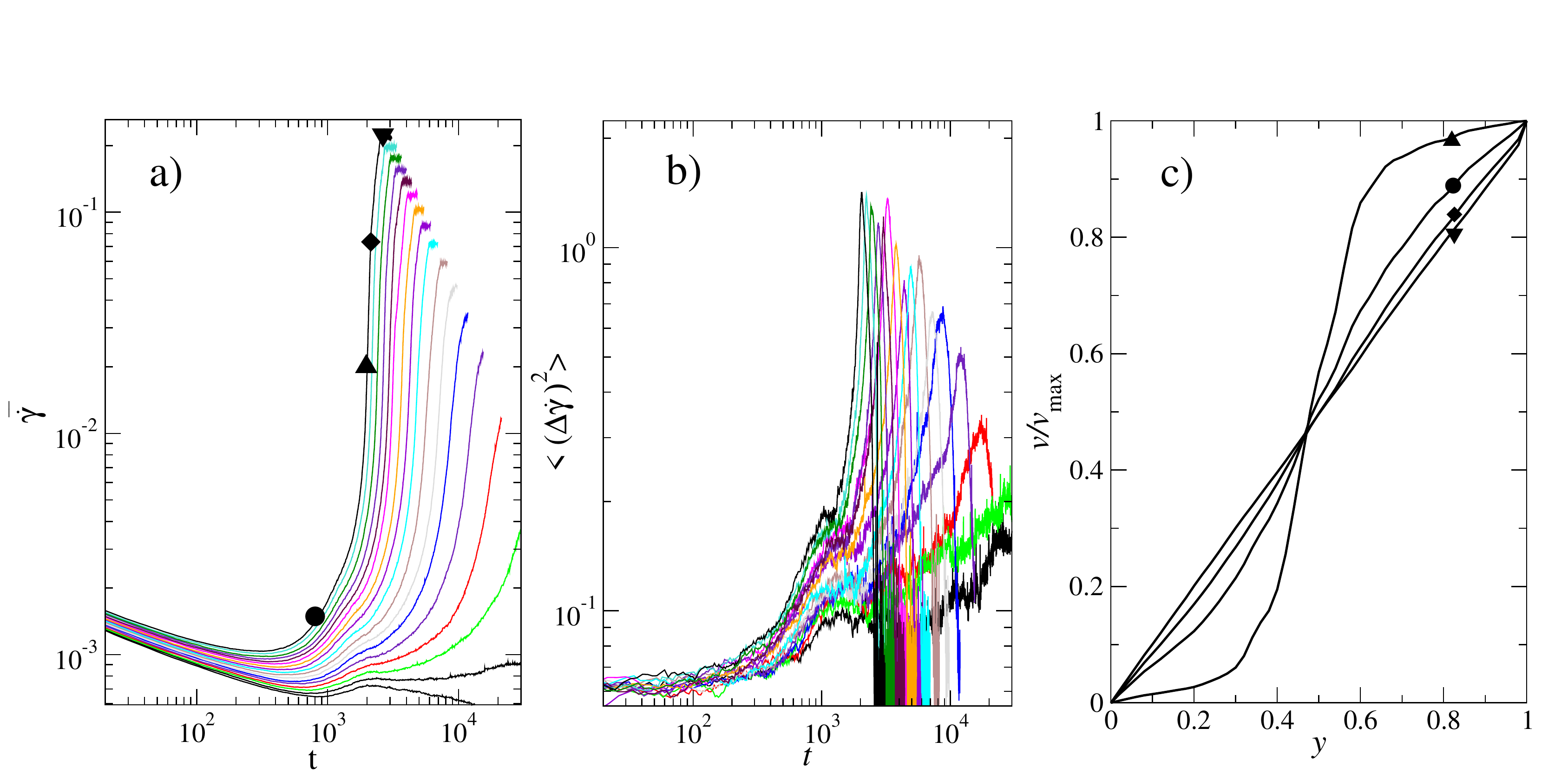}
  \caption{a) Time-differentiated creep curves of the SGR model for
    stress values $\Sigma_0/\Sigma_{\rm y}=1.005,1.010\cdots 1.080$
    (curves upwards).  b) Corresponding degree of shear banding. c)
    Normalised velocity profiles at times denoted by corresponding
    symbols in a) Parameters: $x=0.3, w=0.05,n=50,m=10000$. Initial
    sample age $\tw=10^3\left[1+\epsilon\cos(2\pi y)\right]$,
    $\epsilon=0.1$. Adapted from Fig. 2 of Ref.~\cite{Moorcroftetal2012}.}
\label{fig:SGRcreep}
\end{figure*}
\fi

The relevant rheological response function is then the creep curve
$\gbar(t-\tw,\tw,\Sigma_0)$, which reports the accumulated strain as a
function of the time $t-\tw$ since load application, for any given
waiting time $\tw$ and stress amplitude $\Sigma_0$. In the literature
the results are often in fact instead reported in terms of the
time-differentiated creep (shear rate) curves
$\gdotbar(t-\tw,\tw,\Sigma_0)$, where the overdot denotes
differentiation with respect to $t$.

Experimentally, creep curves are measured by recording the motion of
the rheometer plates relative to each other and so represent the
strain response of the material averaged over the sample as a whole.
In any regime where the deformation remains uniform across the sample,
this measured strain clearly also corresponds to the strain
$\gamma(y)=\gbar$ at each point locally across the gap. Indeed a
commonly made assumption in the literature is that the deformation
will remain uniform.

As surveyed in Sec.~\ref{sec:experiment}, however, many soft glasses
in fact form heterogeneous shear banded states as they creep in
response to an applied shear stress. Motivated by this observation
(and similar ones in polymeric fluids), in
Refs.~\cite{MoorcroftPreprint,Moorcroftetal2012} we performed an
analytical calculation to determine whether (and when) a state of
initially homogeneous creep response might become linearly unstable to
the formation of shear bands. Doing so, we found the criterion for the
onset of banding to be:
\be
\frac{\partial^2\gdotbar}{\partial t^2}/\frac{\partial\gdotbar}{\partial t}>0.
\vspace{0.2cm}
\label{eqn:creepCriterion}
\ee
We argued that this criterion applies universally to all complex
fluids, including the soft glassy materials of interest here.  In this
way, any state of initially homogeneous creep is predicted to become
linearly unstable to the formation of shear bands if its
time-differentiated creep response curve $\gdotbar$ simultaneously
shows upward curvature and upward slope as a function of time. (In
principle it might instead show downward curvature and downward slope,
but we have never in practice seen this in our numerical studies of
soft glasses or polymeric fluids.)

This criterion is derived by means of a linear stability analysis that
considers an underlying base state of initially homogeneous creep
response to the applied load, with strain $\gamma_0(t)$ and strain
rate $\gdot_0(t)$. (For notational convenience we suppress the $\tw$
and $\Sigma_0$ dependencies in writing these quantities here.) To this
base state is added small heterogeneous perturbations to give
$\gdot(y,t)=\gdot_0(t)+\sum_n\delta\gdot_n(t) \exp(in\pi y/L_y)$.
(Other relevant variables in the equations of motion are treated in the
same way, but we do not discuss these details here.) The dynamics of
the perturbations are then studied at linear order in their amplitude.
The regime in which they are found to grow as a function of time,
indicating the onset of shear banding, is found to correspond to that
in which the base state strain rate $\gdot_0$ obeys
(\ref{eqn:creepCriterion}) above.

As such, then, (\ref{eqn:creepCriterion}) technically applies to the
base state shear rate $\gdot_0$ rather than the spatially averaged
signal $\gdotbar$ as measured experimentally by recording the motion
of the rheometer plates.  However $\gdot_0$ and $\gdotbar$ must
clearly coincide in any regime before any significant banding arises.
To determine the onset of banding, therefore,
(\ref{eqn:creepCriterion}) can be applied to the experimentally
measured signal $\gdotbar$ direct.  In this way, bulk rheological data
can be used as a predictor of shear banded flow states, even in the
absence of spatially resolved velocimetry.

Having introduced an onset criterion that we suggest applies to all
complex fluids, we now consider its implications for the soft glasses
of interest here, as modelled by the SGR model in its original form,
which we expect to capture the behaviour of simple YSFs.  As just
argued above, the time-evolution of an underlying homogeneous flow
state can be used to predict the onset of banding.  We shall therefore
first summarise the creep response of the SGR model in its original
form\cite{Moorcroftthesis}, which addresses homogeneous flows only, as
a natural starting point from which to then understand the shear
banding dynamics of the spatially aware
model~\cite{ISI:000266798200007}.

We focus on noise temperatures in the glass phase $x<\xg$, where the
model shows a non-zero yield stress $\Sigmay(x)$. Following the
application of a step stress of amplitude $\Sigma_0<\Sigmay$ below the
yield stress and not too close to it, the system responds by a process
of slow creep with a logarithmically increasing strain $\gamma_0\sim
A(\Sigma_0)\log\left[(t-\tw)/\tw\right]$. The corresponding strain
rate accordingly tends towards zero at long times, $\gdot_0\sim
A(\Sigma_0)(t-\tw)^{-1}$, with the material creeping ever more slowly
as a function of the time since the load was applied.  For stress
values approaching the yield stress from below the prefactor
$A(\Sigma_0)$ becomes very large, with an apparent divergence as
$\Sigma_0\to\Sigmay$ signifying a crossover to a regime in which the
strain no longer behaves logarithmically~\cite{ISI:000085655200008}.
However the strain rate nonetheless still progressively evolves
towards zero at long times.  For applied stress values below the yield
stress, then, the strain rate never satisfies
(\ref{eqn:creepCriterion}) and the creep response is predicted to
remain homogeneous at all times, even in a spatially aware model that
could in principle display banding.

For applied stresses just above the yield stress the system initially
responds in a fashion similar to that for stresses just below yield,
in the sense that it creeps progressively more slowly over time: here
one numerically observes~\cite{Moorcroftthesis} $\gdot_0\sim
\tw^{-1}\left[(t-\tw)/\tw\right]^{-x}$.  However because the applied
stress now exceeds the yield stress, the system must eventually make a
transition to a flowing state with a steady state shear rate
prescribed by the flow curve, $\gdot_0\sim
(\Sigma-\Sigmay)^{1/(1-x)}$. Indeed we find that this transition
occurs via a process of rather sudden fluidisation in which the strain
rate (i) curves upwards to increase from the small value it had
attained by the end of the slow creep regime, then (ii) goes through
an inflexion point, before finally (iii) curving downwards to attain
its ultimate value on the flow curve. (Taking the inflexion point as a
good measure of the fluidisation time, numerically one
finds~\cite{Moorcroftthesis} for the SGR model $\tau_{\rm
  fluidisation}\sim \tw(\Sigma-\Sigmay)^{-\alpha}$ with $\alpha=O(1)$.
As noted above, however, experimentally the actual functional form of
$\tau_{\rm fluidisation}(\Sigma-\Sigmay)$ is found to vary from
material to material.)  Creep curves of this shape have been observed
experimentally in a host of complex fluids, including
mayonnaise~\cite{caton}, hair gel~\cite{caton},
carbopol~\cite{caton,ISI:000294447600069}, a hard sphere colloidal
glass~\cite{siebenbuerger-prl-108--2012}, carbon
black~\cite{sprakel-prl-106--2011} and a lyotropic hexagonal columnar
phase~\cite{bauer-prl-97--2006}.

During part (i) of this fluidisation process the strain rate $\gdot_0$
simultaneously shows upward slope and upward curvature as a function
of time.  It therefore satisfies (\ref{eqn:creepCriterion}), leading
us to predict that the spatially aware version of the model should
become linearly unstable to the onset of shear banding in this regime.
We explored this prediction by performing a full nonlinear simulation
of the spatially aware model~\cite{Moorcroftetal2012}. This
simulation captures not only the initial onset of banding predicted by
the linear instability criterion (\ref{eqn:creepCriterion}), which
applies while any heterogeneous perturbations remain small, but also
nonlinear effects once the heterogeneity becomes significant in the
later stages of band development.

The results are shown in Fig.~\ref{fig:SGRcreep}. The left panel shows
the time evolution of the strain rate signal $\gdotbar$, spatially
averaged across the sample. As argued above, this spatially averaged
quantity must coincide with the counterpart signal $\gdot_0$ of the
homogeneous model in any regime before appreciable banding arises. In
fact our numerics further show it to agree in overall shape even once
banding has set in, following the form described above for $\gdot_0$
throughout its full evolution: a prolonged regime of progressively
slowing creep is followed by a sudden process of fluidisation with
features (i) to (iii) above, with the time of fluidisation apparently
diverging as the applied stress value approaches the yield stress from
above.

In regime (i) of this fluidisation process, in which the shear rate
$\gdotbar$ simultaneously shows upward slope and upward curvature,
significant shear banding indeed arises, consistent with
(\ref{eqn:creepCriterion}). See the middle panel of
Fig.~\ref{fig:SGRcreep}, which shows the evolution of the variance in
shear rate spatially across the gap as a function of time.  This grows
markedly in regime (i), then decays after the inflexion point (ii)
once the shear rate signal curves downward in regime (iii).  Snapshot
velocity profiles at the times indicated by the circles in the left
panel are given in the right hand panel, and indeed exhibit pronounced
shear banding. These predictions of transient banding following
fluidisation after a process of slow creep are consistent with
experimental observations in
Refs.~\cite{ISI:000294447600069,divoux-prl-110--2013,divoux-sm-7-9335-2011},
as discussed in Sec.~\ref{sec:experiment} above.

\subsection{Shear startup protocol: stress overshoot}
\label{sec:startup}

Consider now an experiment in which a sample is freshly prepared at
time $t=0$, left to age in the absence of any applied flow or loading
during a waiting time $\tw$, then for all subsequent times subject to
shear of constant rate $\gdot_0$:
\be
\gdot(t)=\gdot_0\Theta(t-\tw).
\ee
The relevant rheological response function is then the stress startup
curve $\Sigma(t-\tw,\tw,\gdot_0)$ reported as a function of the time
$t-\tw$ since shearing commenced, for a given waiting time $\tw$ and
shear rate $\gdot_0$.  Equivalently one may instead report this as a
function of the accumulated strain $\gamma_0=\gdot_0(t-\tw)$, again
for fixed $\tw$ and $\gdot_0$, to give $\Sigma(\gamma_0,\tw,\gdot_0)$.
For notational convenience in what follows we shall suppress the $\tw$
dependence in writing this, so for any initial sample age $\tw$ we
have $\Sigma(\gamma_0,\gdot_0)$.

\iffigures
\begin{figure*}[tbp]
  \centering
  \includegraphics[width=1.0\textwidth]{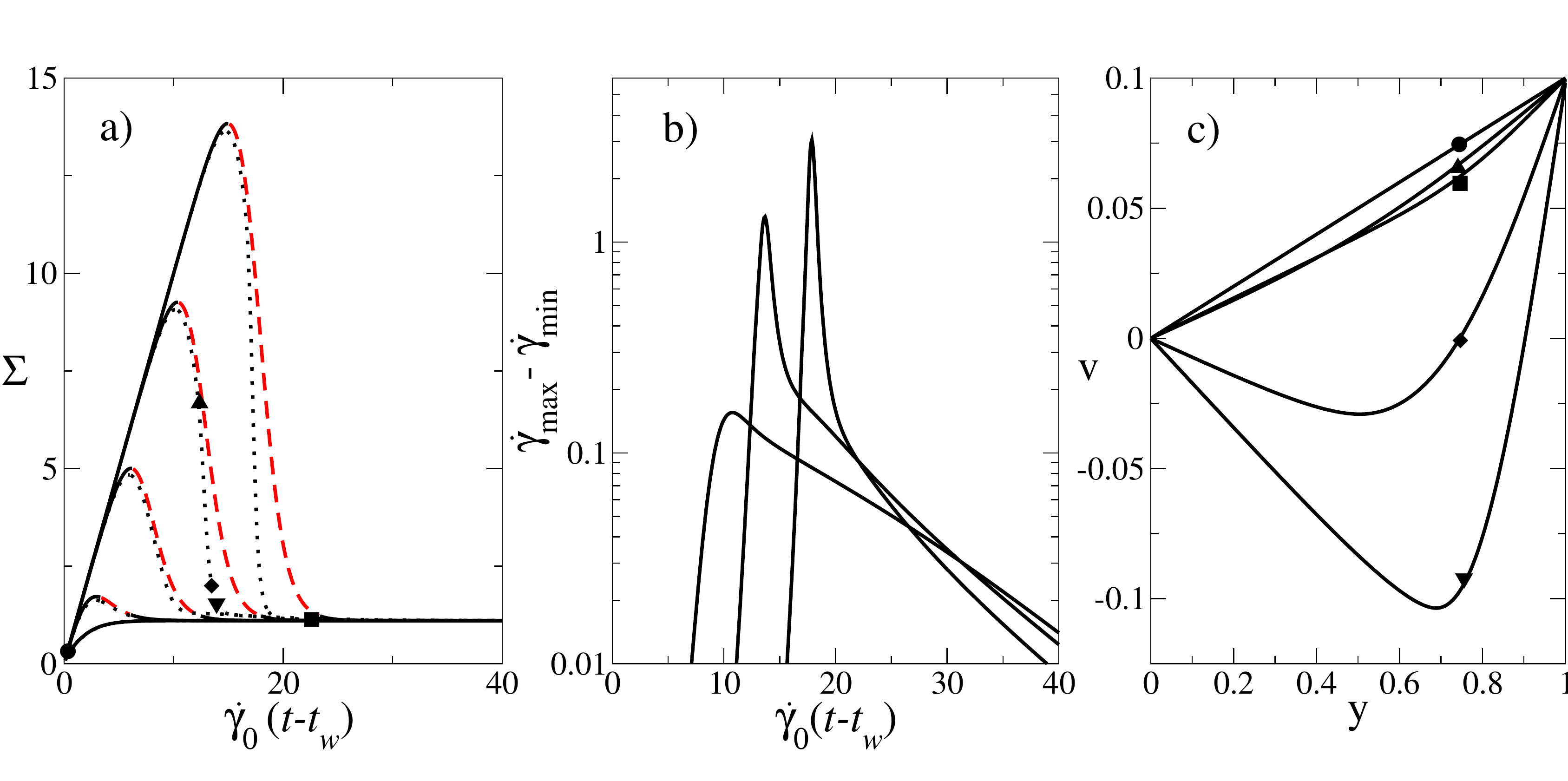}
  \caption{a) Stress startup curves of the fluidity model for an
    imposed shear rate $\gdot_0=0.1$ and different initial sample ages
    $\tw=10^0,10^2,10^4,10^6,10^8$ (curves upwards). Thick lines:
    homogeneously constrained run, with dashed regions denoting
    regimes of linear instability to the onset of banding. Dotted
    lines: full heterogeneous simulation with banding allowed.  b)
    Corresponding degree of shear banding for $\tw=10^4,10^6,10^8$ (in
    order of increasing peak height). c) Velocity profiles at times
    denoted by corresponding symbols in a).  Parameters:
    $\eta=0.005,l_0=0.01,\delta=0.01$. Adapted from Figs. 1 and 2 of Ref.~\cite{ISI:000286879900011}.}
  \label{fig:fluidityStartup}
\end{figure*}
\fi

In the context of shear banding, a familiar thought experiment is then
to consider a startup flow that is (artificially) constrained to
remain homogeneous until a stationary state is attained in the limit
$\gamma_0\to\infty$. In this limit the total accumulated strain
becomes irrelevant, as does the waiting time $\tw$, and the stress
depends only on strain rate, via the underlying homogeneous
constitutive curve
$\Sigma(\gamma_0\to\infty,\gdot_0)=\Sigma(\gdot_0)$. The criterion for
shear banding (with the constraint now removed) is well known in this
limit: that the constitutive curve has a region of negative slope,
$d\Sigma/d\gdot_0<0$. Less artificially, this criterion also marks the
onset of a linear instability to the formation of shear bands in an
experiment in which the shear rate is very slowly swept upwards from
zero.

Because the fluid flows in a liquid-like way in this steady state
limit, we refer to this type of banding, triggered by a regime of
declining stress versus strain rate $d\Sigma/d\gdot_0<0$, as `viscous'
for convenient nomenclature in what follows. As already discussed, the
SGR and fluidity models (at least in their form as introduced above)
each have a monotonic constitutive curve $d\Sigma/d\gdot_0>0$ and so
do not capture steady state viscous banding.

The viscous banding scenario just discussed is analogous to but
distinct from a similar instability known to arise in nonlinear
elastic solids that are subject to an applied shear strain. In this
case, a state of initially homogeneous shear deformation undergoes an
`elastic' instability to the formation of coexisting bands of
differing strain in any regime in which the stress is a declining
function of the applied strain, $d\Sigma/d\gamma_0<0$.

Besides providing an interesting analogy, this elastic instability in
fact has directly important implications for viscoelastic fluids as
well.  Consider a shear startup run performed in the limit of a flow
rate $\gdot_0\to\infty$ that exceeds the fluid's intrinsic
viscoelastic stress relaxation rates.  In this regime, many
viscoelastic materials attain a limiting startup curve
$\Sigma(\gamma_0,\gdot_0\to\infty)=\Sigma(\gamma_0)$ that depends only
on strain, independent of the strain rate, at least while the
accumulated strain remains modest. Once the imposed $\gdot_0$ exceeds
the material's intrinsic relaxation rates, then, performing the run at
any higher $\gdot_0$ would give the same startup curve
$\Sigma(\gamma_0)$.  In this regime, $\gamma_0$ can be thought of as
an elastic strain variable and the material essentially behaves as a
nonlinear elastic solid. If this limiting curve furthermore shows an
overshoot followed by a regime of declining stress
$d\Sigma/d\gamma_0<0$, we expect elastic banding to arise during
startup.

Precisely this scenario was explored in the context of fast shear
startup experiments in polymeric fluids in
Refs.~\cite{MoorcroftPreprint,Moorcroftetal2012}, with an elastic
banding instability being shown to set in around the time of the
stress overshoot.  If the fluid furthermore has negative slope in its
underlying constitutive curve $d\Sigma/d\gdot_0<0$, these elastic
bands can be thought of as the formative precursor of viscous bands
that will persist to steady state $\gamma_0\to\infty$. In contrast,
for a fluid with a monotonic constitutive curve the elastic bands
persist only transiently during startup, eventually decaying to leave
homogeneous shear flow in the final steady state.

Soft glassy materials typically also show a strong stress overshoot in
startup, separating an early time elastic regime in which the stress
grows linearly with the accumulated strain $\Sigma=G\gamma_0$, from a
final steady state in which the stress is prescribed by a balance
between elastic loading and plastic relaxation. In these materials,
however, the overshoot cannot be attributed a purely elastic origin
because the decrease in stress post-overshoot arises directly from the
onset of the plastic relaxation processes that lead to the eventual
steady flowing state.  Accordingly, $\gamma_0$ is not an elastic
strain variable in the regime where $d\Sigma/d\gamma_0<0$, and there
can be no direct mapping to an elastic banding scenario.  Soft glasses
have nonetheless been shown to exhibit pronounced banding in startup,
which furthermore does appear closely associated with stress
overshoot, as surveyed in Sec.~\ref{sec:experiment} above. 
Motivated by these observations, we now discuss shear startup
in the fluidity and SGR models, following
Ref.~\cite{ISI:000286879900011}.

As already discussed, the onset of banding in a time-dependent flow
protocol can be predicted by considering an underlying time-evolving
base state of homogeneous shear response to the imposed deformation,
then studying the dynamics of heterogeneous perturbations to this base
state (both at the level of linear instability while the perturbations
remain small, then nonlinear dynamics once noticeable bands have
developed later on).  In what follows, therefore, we shall first
consider the stress startup curves of the fluidity and SGR models with
an artificially imposed constraint of homogeneous shear in each case.
(As noted above these must also coincide with the startup curves
measured experimentally, at least until any significant banding
arises.) We shall then demonstrate that the presence of an overshoot
in the startup curve is closely associated with the onset of an
instability to the formation of shear bands in a model that does allow
spatial variations, and accordingly also experimentally.

Startup curves $\Sigma(\gamma_0,\gdot_0)$ of the fluidity model are
shown in Fig.~\ref{fig:fluidityStartup}a) for several different sample
ages $\tw$, for a base state flow that is constrained to remain
homogeneous. These display an overshoot that depends strongly on the
sample age $\tw$, occurring at a strain $\gamma_0=\log(\gdot_0\tw)$
(to within logarithmic corrections) and a corresponding stress
$\Sigma_0\approx G\log(\gdot_0\tw)$. (Experimentally, however, the
functional form is often observed to be a weak power
law~\cite{divoux-sm-7-9335-2011,mann7}.) As discussed above, the
presence of an overshoot followed by a regime of declining stress
$d\Sigma/d\gamma_0<0$ is expected to give rise to the formation of
shear bands. A linear stability analysis of the dynamics of small
heterogeneous perturbations about the evolving homogeneous base state
indeed confirms this, with a regime of instability indicated by the
red dashed lines Fig.~\ref{fig:fluidityStartup}a.

As can be seen, the regime of instability is much more pronounced for
larger values of the waiting time $\tw$, consistent with the degree of
instability being controlled by the size of the overshoot, which is
much larger in samples that were first aged into a more elastic state
before the flow commenced.  Accordingly, we expect much more
pronounced transient banding to arise in samples that are left for a
long time after preparation before the startup of flow.  In young
samples, in contrast, any region of instability will be sufficiently
weak and short lived that no observable banding can develop during
startup.

With these considerations in mind, we now turn to the full
heterogeneous dynamics of the fluidity model, performing nonlinear
simulations that also allow for spatial variations in the flow
gradient direction $y$.  Each run is initialised with a small
perturbation $\sigma(y,t=0)=\delta\cos(\pi y)$ with $\delta\ll 1$, in
order to seed any banding. As a function of shearing time $t-\tw$ (or
equivalently of accumulated strain $\gdot_0(t-\tw)$) we then track the
degree of shear banding across the sample, which we measure at any
instant by the difference $\gdot_{\rm max}-\gdot_{\rm min}$ between
the maximum and minimum shear rates present in the cell. The evolution
of this quantity is shown in Fig.~\ref{fig:fluidityStartup}b.  Regimes
of high $\gdot_{\rm max}-\gdot_{\rm min}$ indeed match up with those
of negative slope in the startup curves, and with much more pronounced
transient banding for the older samples, as anticipated above.

Snapshots of the flow state at representative times during one run are
shown in Fig.~\ref{fig:fluidityStartup}, and indeed exhibit pronounced
shear banding.  Note that the shear rate in the low shear rate band in
fact becomes negative during startup, consistent with this band
responding essentially like an elastic slab subject to a declining
stress post-overshoot: an elastic material being unloaded will indeed
shear backwards. At longer times the bands decay to leave the flow
homogeneous in the final steady state, consistent with the underlying
constitutive curve of the model being monotonic in a simple YSF,
$d\Sigma/d\gdot_0>0$.

This transient formation of shear bands also slightly perturbs the
startup curves in Fig.~\ref{fig:fluidityStartup}a, leading to a reduced stress
relative to that of the homogeneous startup flow. The overall shape of
the curves is however qualitatively unaffected. This is consistent
with our claim made above, that bulk rheological data can be used as
predictor of the presence of shear banding within the sample, even in
the absence of spatially resolved velocimetry.

The same overall behaviour is seen in the SGR model, with an
age-dependent stress overshoot triggering the formation of pronounced
transient shear banding during startup (not shown).  The same was also
seen in a model of shear transformation
zones~\cite{manning-pre-76--2007} and in an elasto-plastic
model~\cite{jagla-jsme---2010}. The observation of the same behaviour
in four different models leads us to suggest that shear banding
associated with startup overshoot must arise generally in soft glassy
materials that have been aged prior to shear.

\iffigures
\begin{figure*}[tbp]
  \centering
\includegraphics[width=1.0\textwidth]{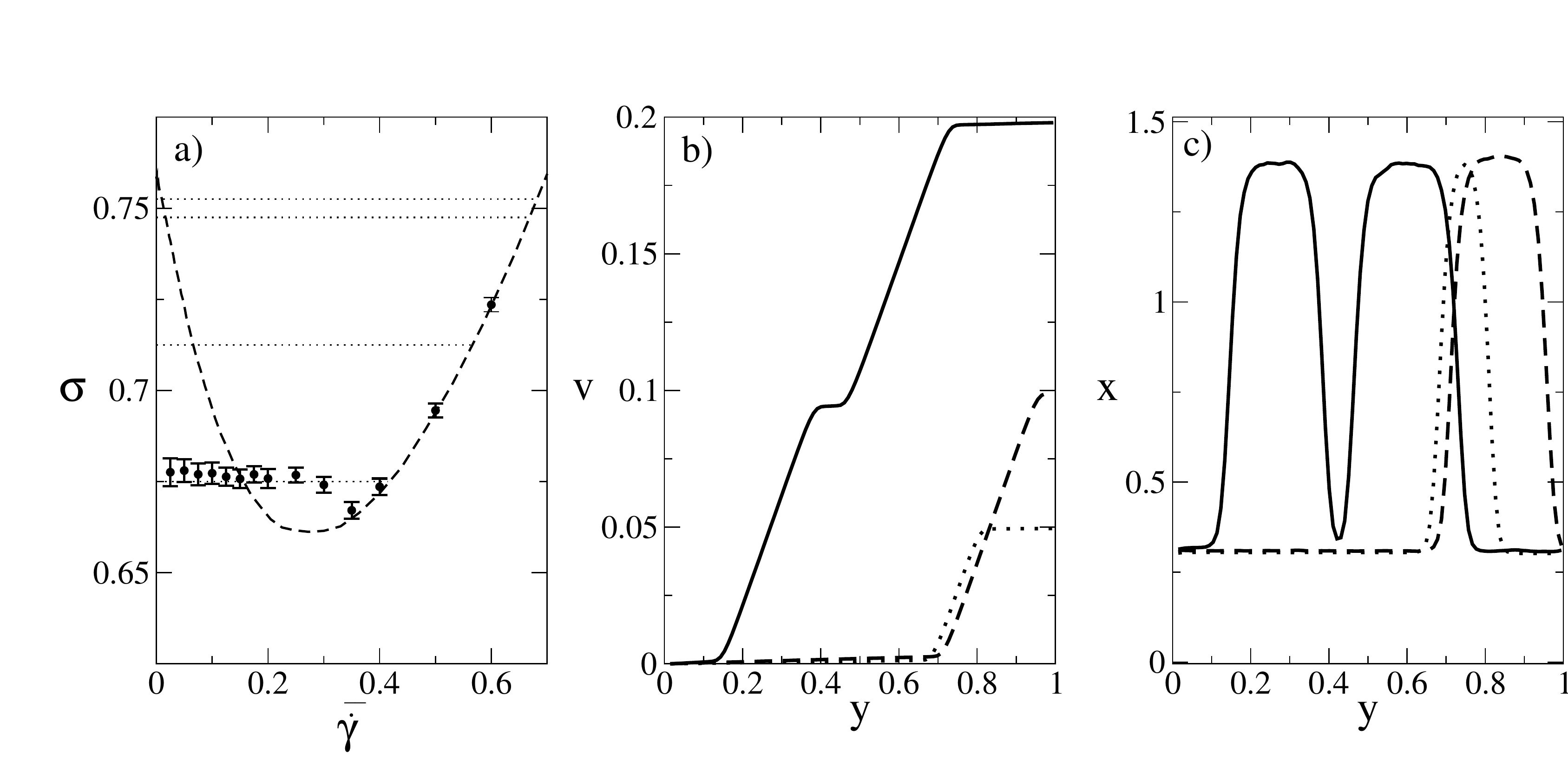}
\caption{a) Dashed line: constitutive curve of stress against strain
  rate for homogeneous shear states of model 1 for $x_0=0.3$,
  $a=2.0$. Symbols: stress at various mean imposed shear rates
  $\gdotbar$ found in the long time limit after startup of steady
  shear. Shear banding is present in any regime where the symbols
  differ significantly from the constitutive curve. b) Velocity
  profiles for imposed mean shear rates $\gdotbar=0.05,0.1,0.2$
  (dotted, dashed, solid). c) Corresponding profiles of noise
  temperature. Adapted from Figs. 1 and 2. of
  Ref.~\cite{ISI:000266798200007}.}
  \label{fig:model1}
\end{figure*}
\fi

There is however a new feature in the SGR model, not seen in the
fluidity model.  For the oldest samples the time scale for the bands
to decay back to a homogeneous flow can be inordinately long,
requiring thousands of strain units. (Not shown; see
Ref.~\cite{ISI:000286879900011} for details.)  Indeed such strains may
be unattainable in any realistic experiment, in which case shear
banding would represent the ultimate flow response of the material for
practical purposes, even though the underlying constitutive curve of
the material is monotonic and any true steady state homogeneous.
(Because strain in general rejuvenates a soft glass, this long
lifespan of the low shear band is possible only because the strain
rate in it remains extremely small compared to the average shear rate
applied to the sample as a whole.) We return in
Sec.~\ref{sec:conclusion} below to discuss the implications of such
long-lived bands, even with a monotonic constitutive curve, for the
apparent distinction between simple YSFs and viscosity-bifurcating
YSFs that is widely discussed in the experimental literature.

As noted above, in soft glasses the presence of a stress overshoot
during shear startup arises from a competition between elastic loading
and plastic relaxation. In consequence there can be no direct mapping
either to a purely elastic instability, $d\Sigma/d\gamma_0<0$, or a
purely viscous instability, $d\Sigma/d\gdot_0<0$: the instability
reported here represents an interesting new intermediate between these
two limiting cases.

The predictions reported in this section are consistent with
observations of transient shear banding triggered by stress overshoot
during shear startup in a simple YSF in
Refs.~\cite{divoux-sm-7-9335-2011,divoux-prl-104--2010}, which decays to
leave homogeneous flow in steady state. In fact our predictions for an
initial onset of shear banding triggered by stress overshoot are also
consistent with data in viscosity-bifurcating YSF in
Ref.~\cite{martin-sm-8-6940-2012}, and in that case the bands persists
permanently, as long as the flow remains applied, consistent with the
more complex rheology of viscosity-bifurcating YSFs. It is to these
permanent bands in viscosity-bifurcating YSFs that we now turn.

\section{Viscosity-bifurcating yield stress fluids}
\label{sec:permanent}

In its form as described so far, the SGR model has a monotonic
underlying constitutive curve $\Sigma(\gdot)$ and does not admit
permanent banding under conditions of a steady applied shear flow.
This apparently contradicts experimental observations in
viscosity-bifurcating
YSFs~\cite{PhysRevLett.88.175501,martin-sm-8-6940-2012,paredes-jpm-23--2011,rogers-prl-100--2008,RT114}
(though see comments in the closing section~\ref{sec:conclusion}
below).
%
This failure of the SGR model to admit non-monotonic constitutive
curves
may be linked to the fact that its noise temperature $x$ is
taken to be a constant parameter of the model. In practice, however,
$x$ is not the true temperature but represents in a mean field way
coupling between yield events occurring in different parts of the
sample. As such, the effective noise temperature experienced by any
given element should in fact depend on the level of hopping activity
within that element's local vicinity.

With this physical picture in mind, we now move beyond our assumption
of a constant $x$ to consider the following picture of
relaxation-diffusion dynamics:
\be
\label{eqn:x}
\tau_x\dot{x}(y,t)=-x+x_0+S+\lambda^2\frac{\partial^2x}{\partial
  y^2}=0.  
\ee
In the second equality we have for simplicity set $\tau_x\to 0$ so
that the noise temperature rapidly adapts to changes in nearby
activity levels.  

The diffusive term in this equation obviates the need for any
stochastic diffusive dynamics of the kind discussed at the end of
Sec.~\ref{sec:SGR} above, so we henceforth set the parameter $w$
introduced in that section to zero.  Besides these changes, the model
is otherwise unchanged from the dynamics defined previously.

The source term $S(y)$ in Eqn.~\ref{eqn:x} represents pumping of the
noise by hopping events, and so depends on the probability
distribution $P(E,l,y,t)$ at position $y$. We henceforth denote this
$P(y)$ for notational convenience. In what follows we shall explore
two model variants, based on different choices for this source term.

\begin{itemize}

\item Model 1 has
\be
\label{eqn:source1}
S(y)=a\langle l^2/\tau\rangle_{P(y)},
\ee
where $\tau=\tau_0\exp\left[(E-kl^2/2)/x\right]$ is the trap lifetime.
Within this choice, the noise is assumed to be pumped by the
dissipation of elastic energy.

\item Model 2 instead has
\be
\label{eqn:source2}
S(y)=\tilde{a}\langle 1/\tau\rangle_{P(y)},
\ee
in which all hops contribute equally to the noise, regardless of the
local strain released in any hop.
\end{itemize}

We first explore our results for model 1. In a homogeneous steady
state, Eqns.~\ref{eqn:x} and~\ref{eqn:source1} are together equivalent
to
\be
x=x_0+2a\Sigma(x,\gdot)\gdot.
\ee
This implicit relation allows us to construct the homogeneous
constitutive curve $\Sigma(\gdot)$, for any $x_0$ and $a$, as a
composite combination of the constant-$x$ curves of the original model
in Fig.~\ref{fig:SGRconstitutive}. See the dashed line in
Fig.~\ref{fig:model1}a). As can be seen, the constitutive curve now
has non-zero yield stress $\Sigma_y$ in the limit $\gdot\to 0$,
followed by a region of declining stress $d\Sigma/d\gdot<0$, before
restabilising at higher shear rates. In a step stress protocol, this
gives rise to a classic viscosity-bifurcation scenario: see Fig.~3 of
Ref.~\cite{ISI:000266798200007}.

\iffigures
\begin{figure}[tbp]
  \centering
  \includegraphics[width=0.5\textwidth]{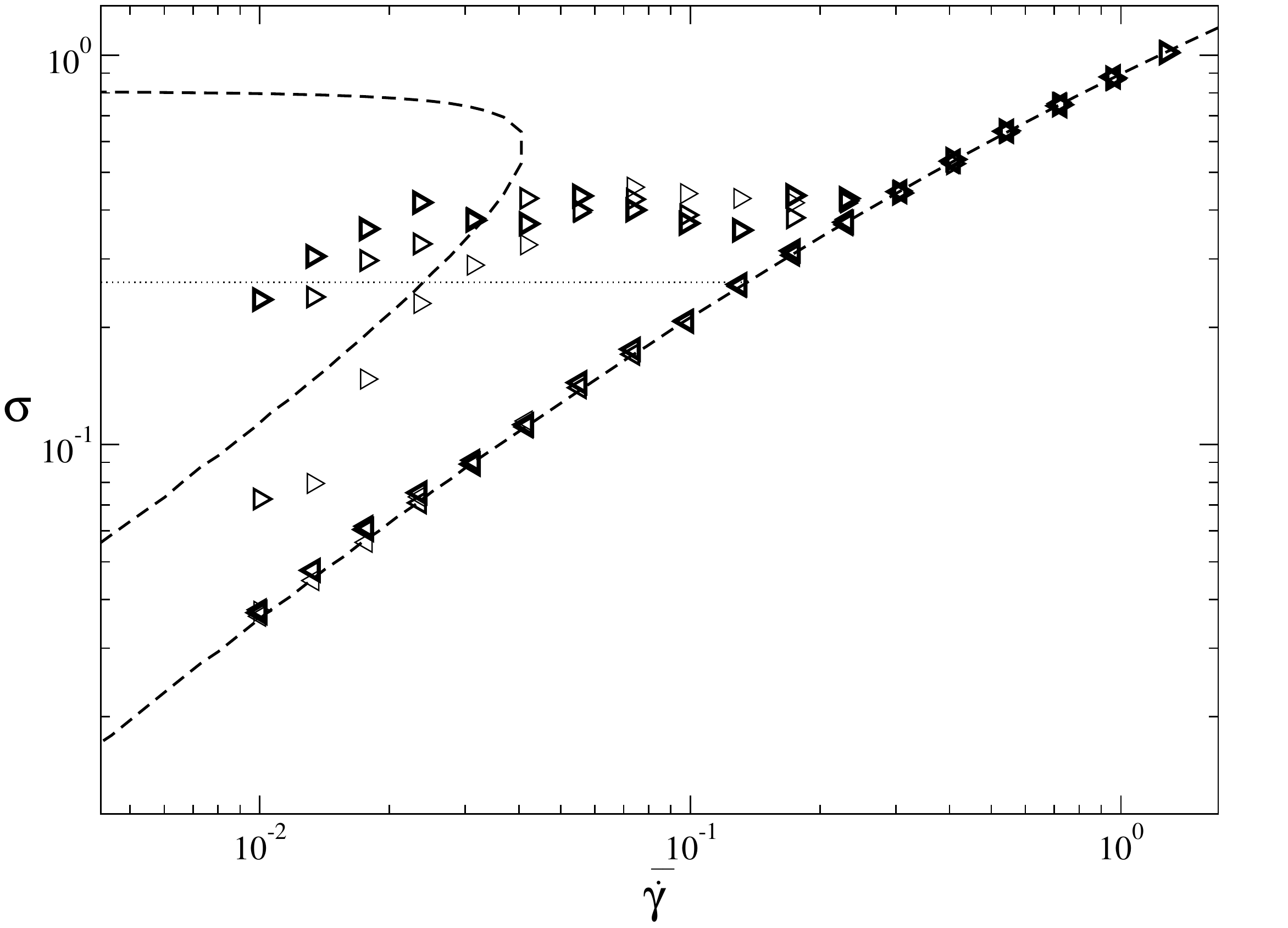}
  \caption{Dashed lines: constitutive curve of shear stress against
    shear rate for homogeneous flow states in model 2 with
    $x_0=0.15,\tilde a=3.75$. Symbols: stress values in spatially
    resolved waiting time Monte Carlo simulations ($n=100, m = 1000,
    \lambda = 0.5\Delta$) for up/down strain-rate sweeps
    (right-pointing and left-pointing triangles respectively) with a
    residence time per point of $t_r = 200,400,800$ (thin, medium,
    bold symbols). In each run the system was initialized in a
    homogeneously aged state of $t_w = 10^4$.  Dotted line shows, as a
    guided to the eye, the quasi-steady stress attained at long times
    in shear startup for $\overline{\dot\gamma} \le 0.1$ in the shear
    banding regime. Reproduced from Fig. 4 of Ref.~\cite{ISI:000266798200007}.}
  \label{fig:model2}
\end{figure}
\fi

This nonmonotonicity also creates the standard preconditions for
permanent shear banding under conditions of a constant imposed shear
rate. Unusually, however, compared with more familiar ergodic shear
banding fluids such as wormlike micelles, the presence of a yield
stress in Fig.~\ref{fig:model1} implies that the viscous band will be
effectively unsheared.  A waiting time Monte Carlo simulation of the
model's full spatio-temporal dynamics confirms this scenario: see
Fig.~\ref{fig:model1}.  The low shear band is indeed effectively
solid, with a strain rate close to zero and a low noise temperature
$x\approx x_0=0.3$ such that $\Sigmay(x) > \Sigma$.  Under these
conditions, ergodicity is broken in this low shear band and the
dynamical correlator $C(t, \tw)$, which measures the fraction of
unhopped particles, exhibits simple aging when measured locally in
this band (not shown).  In contrast, the high-shear band has a high
level of activity that self-consistently maintains it in an ergodic
state of high $x$ and low viscosity.  Plotting the stress as a
function of the overall imposed shear rate $\gdotbar$ in this shear
banding regime then gives a plateau in the flow curve that is strongly
characteristic of shear banded flows.  We further find the relative
volume fractions of the bands to obey a lever rule, as seen
experimentally~\cite{RT114}.

Other theoretical approaches to invoke a non-monotonic constitutive
curve and capture permanent shear banding in steady shear include
fluidity models besides the one used
here~\cite{picard-pre-66--2002,coussot-jr-46-573-2002}, models of
coupled elastic plastic
events~\cite{mansard-sm-7-5524-2011,martens-sm-8-4197-2012,jagla-pre-76--2007}
and STZ theories~\cite{manning-pre-79--2009}. These findings, and our
predictions just discussed, are consistent with experimental reports
in viscosity-bifurcating
YSFs~\cite{PhysRevLett.88.175501,martin-sm-8-6940-2012,paredes-jpm-23--2011,RT114}.

We next turn to Model 2, which, as we shall show, has constitutive
curves of a shape that allows us to address recent experiments on star
polymers~\cite{rogers-prl-100--2008,rogers-jr-54-133-2010}. Under
conditions of a relatively rapid upward shear-rate sweep (with a
residence time $t_{\rm r}= 10 s$ per observation point), these
materials experimentally show an apparently conventional monotonic
flow curve.  In contrast, slower sweeps with $t_{\rm r} = 10^4 s$ give
a much larger stress that is almost constant at small values of the
imposed shear rate.  NMR velocimetry reveals the presence of shear
banding in this regime, with the viscous band effectively unsheared,
$\gdot=0$. A strong hysteresis is also seen, with the less viscous
branch persisting to much lower strain rates on sweeping the applied
shear rate down towards zero again.

As can be seen in Fig.~\ref{fig:model2}, this experimental scenario is
indeed captured by model 2. The right-pointing triangles show the
model's stress response to a slow upward shear rate sweep for a sample
of age $\tw = 10^4$ before shear.  This exhibits an obvious stress
plateau for shear rates $\gdot < 0.3$, which is the signature of
coexisting glassy and flowing shear bands (not shown).  (At the lowest
applied shear rates, the stress does not have time to attain this
plateau stress before the strain rate is swept on to a higher value.)
For shear rates $\gdot > 0.3$ the system flows homogeneously on the
fluid branch of the constitutive curve. A remarkable feature of Model
2, not seen in Model 1, is that the constitutive curve $\Sigma(\gdot)$
is multivalued, in both stress and strain rate, down to the lowest
accessible shear rates. In consequence, after the sample has been in a
flowing state on the fluid branch at a high value of $\gdot$, it can
remain in a homogeneous fluidized state even as the shear rate is
ramped back down to zero: see the left-pointing triangles in
Fig.~\ref{fig:model2}. This is consistent with the experiments on star
polymers in Refs.~\cite{rogers-jr-54-133-2010,rogers-prl-100--2008}.

In this section on viscosity bifurcating YSFs, we have focussed mainly
on the permanent shear banding effects that are unique to this class
of fluids, and not seen in simple YSFs. Nonetheless, the
time-dependent banding effects reported in Sec.~\ref{sec:transient}
above for simple YSFs are expected to arise in viscosity-bifurcating
YSFs as well. In particular, we anticipate band formation triggered by
stress overshoot during shear startup, and now persisting to steady
state for shear rates in the plateau regime of the flow
curve~\cite{martin-sm-8-6940-2012}.  We also expect transient shear
banding associated with sudden fluidisation under conditions of an
imposed step stress in the vicinity of $\Sigmay$.

\iffigures
\begin{figure}[tbp]
  \centering
  \includegraphics[width=0.5\textwidth]{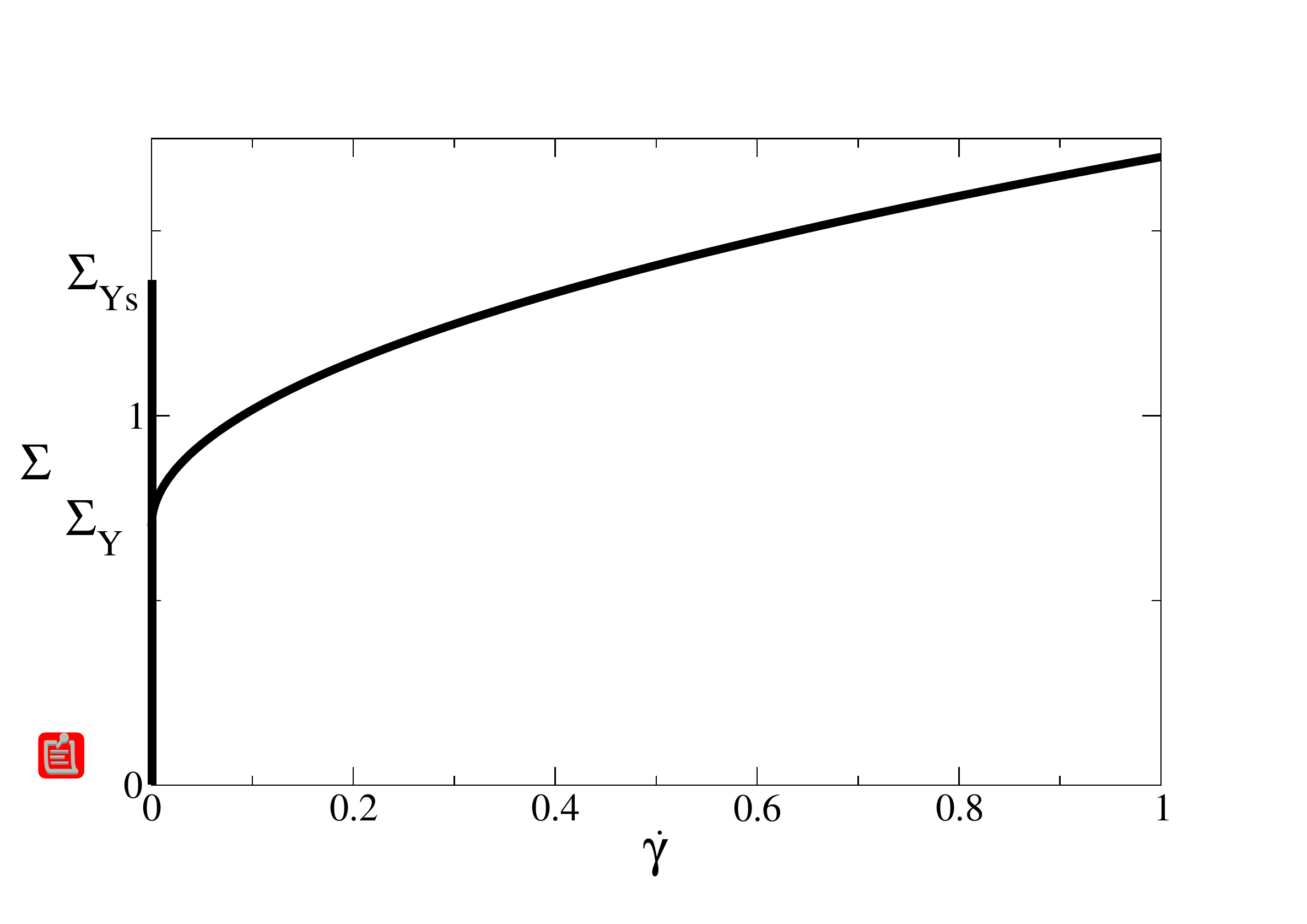}
  \caption{Sketch of a static yield stress exceeding the dynamic one,
    allowing a possible coexistence of static and flow shear bands at
    a common value of the stress.
  \label{fig:sketch}}
\end{figure}
\fi

\section{Conclusions  and outlook}
\label{sec:conclusion}

In this paper, we have reviewed recent theoretical progress in
addressing widespread observations of shear banding in soft glassy
materials. Following introductory remarks in Sec.~\ref{sec:intro}, we
started in Sec.~\ref{sec:experiment} by surveying the experimental
phenomenology, considering in particular a distinction that is widely
discussed in the literature: between viscosity-bifurcating yield
stress fluids (YSFs), which apparently display permanent shear banding
under conditions of a steady applied shear flow, and simple YSFs, in
which shear bands arise only transiently during the process whereby a
steady flowing state is established out of an initial rest state. (We
also noted that viscosity-bifurcating YSFs are often referred to as
thixotropic YSFs in the literature, but cautioned against this
nomenclature on account of the fact that simple YSFs can also display
ageing and rejuvenation effects -- {\it i.e.}, thixotropy -- in their
time-dependent rheology.)

In Sec.~\ref{sec:models} we introduced the models to be used
throughout the paper: the soft glassy rheology (SGR) model and a
simple fluidity model. In original form, these both have a monotonic
constitutive relation between shear stress and shear rate (for an
underlying base state of homogeneous shear flow) and are therefore
unable to capture permanent shear banding under conditions of a steady
applied shear flow.  They do nonetheless convincingly capture
observations of shear banding in time-dependent flow protocols such as
shear startup and step stress: recall Sec.~\ref{sec:transient}.
Although technically only transient, these bands may persist for
several hundreds or even thousands of strain units and so represent
the ultimate flow response of the material for practical purposes,
even in a simple YSF for which the true steady state would in
principle be unbanded.

We also discussed these results in the broader context of recently
predicted criteria for the onset of shear banding in time-dependent
flow protocols~\cite{Moorcroftetal2012}. (These criteria in fact apply
to all complex fluids, and not just the soft glassy materials of
interest here. A detailed investigation of the directly analogous
phenomena in ergodic complex fluids such as polymers and wormlike
micellar surfactants can be found in Ref.~\cite{MoorcroftPreprint}.)
An important prediction of this work is that shear banding should
arise generically in any system where stress overshoots arise in shear
startup, and where the time-differentiated creep curve shows a regime
of upward curvature following the imposition of a step stress.
Eventually, one may also wish to consider other time-dependent
protocols, besides the case of shear startup and step stress reported
here.

With regards flow heterogeneity triggered by upward curvature in the
time-differentiated creep curve following the imposition of a step
stress, is is interesting to note that two separate regions of upward
curvature are observed experimentally in Fig. 1 of
Ref.~\cite{ISI:000294447600069}: the first corresponding to a regime
of total wall-slip and the second to transient shear banding. Whether
total wall-slip could be decsribed in terms of a shear banding-like
instability near the walls, and how to incorporate wall effects into
SGR and fluidity models, remains an open question.

To address the possibility of permanent shear banding under conditions
of a steady applied shear flow, as seen in viscosity-bifurcating YSFs,
we introduced in Sec.~\ref{sec:permanent} a simple variant of the SGR
model in which the noise temperature responds dynamically to the local
rate of activity.  This captures a non-monotonic constitutive curve
and so allows permanent banding in a steady applied shear flow.  A
non-monotonic variant of the fluidity model does likewise, as explored
in Refs.~\cite{picard-pre-66--2002,coussot-jr-46-573-2002}.

As seen in Fig.~\ref{fig:model1}, the form of the non-monotonicity (in
model variant 1) comprises a branch of zero-flow states for stresses
$\Sigma<\Sigmay$, followed by a branch of flow states in which the
stress first decreases with strain rate before rising again in faster
flows.  Molecular dynamics
studies~\cite{varnik-prl-90--2003,varnik-jcp-120-2788-2004,xu-pre-73--2006,chaudhuri-pre-85--2012}
elsewhere in the literature suggest instead the existence of a static
yield stress $\Sigma_{\rm Ys}$ below which there exists (for a
previously unsheared sample) a branch of zero flow states, with
$\Sigma_{\rm Ys}$ exceeding the dynamical yield stress $\Sigma_{\rm
  Y}$ as measured in a protocol that instead sweeps the shear rate
down towards zero.  (That dynamical branch of the constitutive curve
is itself purely monotonic.)  This gives rise to a downwards step
discontinuity (non-monotonicity) in the material's constitutive
properties at $\gdot=0$, allowing the coexistence of an unsheared band
with a flowing one. See Fig.~\ref{fig:sketch}. 

In Ref.~\cite{chaudhuri-pre-85--2012} it was suggested that this
difference between $\Sigma_{\rm Ys}$ and $\Sigma_{\rm Y}$, which gives
rise to the discontinuity just described, may persist only for a
finite (but very long) duration, and only in a system of finite size
(disappearing for an infinitely large system).  Indeed, the shear
bands observed in that study were very long lived, but not a true
zero-frequency phenomenon. Indeed this scenario may not be entirely
distinct from the observation of extremely long lived (though
technically transient) shear bands in the original SGR model.

Taken together, these considerations raise the intriguing possibility
that the apparent difference between a viscosity-bifurcating YSF and a
simple YSF might in some cases lie not in a true zero-frequency
difference in the structure of their constitutive curves, but rather
in the presence of inordinately long (not nonetheless still finite)
timescales in viscosity-bifurcating YSFs, which in simple YSFs are
instead merely shorter and more commensurate with experimental
timescales.  Indeed, dialing progressively more attractive particle
interactions into the simulations Ref.~\cite{chaudhuri-pre-85--2012}
appeared to increase the window of shear rates affected by
$\Sigma_{\rm Ys}$, leading to correspondingly more pronounced and
longer lived bands. This question of true non-monotonicity versus
transient apparent non-monotonicity clearly deserves further careful
thought.

Transient shear banding during time-dependent shear in a polymer glass
was recently reported theoretically in Ref.~\cite{ron}. Clearly, it
would be interesting to explore in more detail spatially heterogeneous
deformation of polymer glasses both theoretically and experimentally.

Although the primary focus of this article has not been hard sphere
colloidal glasses, a few remarks concerning the potential relevance of
the phenomena reviewed here to that class of materials are worthwhile.
Overarching this review has been the concept that particular
rheological signatures in an underlying base state of homogeneous flow
can signify a linear instability of that base state, which leads to
the formation of a heterogeneous, shear banded flow. In terms of a
material's long-time response, that base-state signature is typically
that the underlying constitutive curve of stress as a function of
strain rate $\Sigma(\gdot)$ is non-monotonic, leading to steady state
shear banding. Another possible mechanism for true steady state shear
banding {\em without} a non-monotonic constitutive curve is that of
coupling between flow and concentration fluctuations, via normal
stresses. (This has long been known to arise in polymer and wormlike
micellar surfactant solutions: see~\cite{EPJE} and references
therein.)  We have not discussed that mechanism in detail here,
because the SGR model does not allow for concentration
fluctuations. However, precisely that mechanism was explored in the
context of steady shear banded states of hard sphere colloidal glasses
in Ref.~\cite{ISI:000286754700021}.  And besides the possibility of
steady state banding, in a time-dependent shear startup flow the
signature of a non-monotonic stress versus strain $\Sigma(\gamma)$ in
an initially homogeneous base state can signify an instability leading
to the formation of shear bands during startup. This was demonstrated
in the SGR and fluidity models above, and has widely been seen
experimentally - recall Sec.~\ref{sec:experiment}. Note that this
overshoot-driven instability is an {\em additional} mechanism that can
cause shear banding, beyond the two mechanisms for steady state bands
discussed above. These `startup' bands may or may not then persist to
steady state, depending whether either of the steady-state mechanisms
(a non-monotonic $\Sigma(\gdot)$, or alternatively  an unstable
coupling between flow and concentration) is present. Because many
existing theoretical studies of hard sphere glasses that impose a
homogeneous shear flow, including mode coupling
theories~\cite{Laurati2012,Amann2013} and Brownian
simulations~\cite{Laurati2012,Koumakis2012}, predict strong stress
overshoots in shear startup, it would clearly be very interesting to
perform the counterpart heterogeneous calculations and simulations to
check for the possiblity of shear banding in startup.  It could also
be worth checking for shear bands that arise transiently during
startup due to stress overshoot (or that may intially set in during
startup then persist to steady state because of coupling to
concentration) in experiments on colloidal
glasses~\cite{Amann2013,siebenbuerger-prl-108--2012,Koumakis2012}.

Finally, an effect that arises virtually ubiquitously in the rheology
of complex fluids is that of wall slip: reports are widespread in the
experimental literature, ranging from the anecdotal to the carefully
considered. The drastic effect of boundary conditions and wall slip on
yielding phenomena was reported in particularly careful detail in
Refs.~\cite{gibaud-prl-101--2008,requested6}.  Clearly, any
manifestations of wall slip become increasingly more important in
confined geometries, as do any non-local effects in the fluid's
rheological
properties~\cite{goyon-n-454-84-2008,mansard-sm-8-4025-2012}. Detailed
theoretical studies of both wall slip (which is unaccounted for by the
SGR and fluidity models in their present form) and non-locality remain
open challenges for the future.

\section{Acknowledgements}

The author thanks Mike Cates, Robyn Cooke (n\'{e}e Moorcroft) and
Peter Sollich for helpful discussions and enjoyable collaboration on
these topics. The research leading to these results has received
funding from the European Research Council under the European Union's
Seventh Framework Programme (FP7/2007- 2013) / ERC Grant agreement no.
279365; and from the UK's EPSRC (EP/E5336X/1).

The author dedicates this manuscript to the memory of Professor Sir
Paul Callaghan.

\bibliographystyle{prsty} 



\end{document}